\documentclass[usenatbib]{mn2e}
\usepackage{amssymb}
\usepackage{times}
\usepackage{graphicx}
\usepackage{epsfig}
\usepackage{rotating}

\newcommand{\beq}{\begin{eqnarray}}
\newcommand{\eeq}{\end{eqnarray}}
\newcommand{\be}{\begin{equation}}
\newcommand{\ee}{\end{equation}}

\newcommand{\xte}{{\textit{RXTE}}}

\newcommand{\g}{$\gamma$}
\newcommand{\phiorb}{\phi}
\newcommand{\phisup}{\Phi}
\newcommand{\psup}{P_{\rm sup}}
\newcommand{\orbshift}{\phiorb_{\rm b}}
\newcommand{\supshift}{\Delta\phisup}
\newcommand{\rb}{r_{\rm b}}
\newcommand{\betaj}{\beta_{\rm j}}
\newcommand{\gammaj}{\gamma_{\rm j}}
\newcommand{\tauw}{\tau_{\rm iso,0}}
\newcommand{\thetamax}{{\theta_{\max}}}
\newcommand{\etafw}{\eta_{\rm fw}}

\title[Superorbital variability of Cyg X-1]{Superorbital variability of X-ray and radio emission of Cyg X-1 --\\ 
II. Dependence of the orbital modulation and spectral hardness on the superorbital phase}

\author[J. Poutanen, A. A. Zdziarski and A. Ibragimov]
{Juri Poutanen,$^1$\thanks{E-mail: juri.poutanen@oulu.fi (JP), aaz@camk.edu.pl (AAZ), askar.ibragimov@oulu.fi (AI)}
Andrzej A. Zdziarski$^2$\footnotemark[1] and Askar Ibragimov$^{1,3}$\footnotemark[1]\\
$^1$Astronomy Division, Department of Physical Sciences, PO Box 3000, FIN-90014 University of Oulu, Finland\\
$^2$Centrum Astronomiczne im.\ M. Kopernika, Bartycka 18, 00-716 Warszawa, Poland\\
$^3$Kazan State University, Astronomy Department, Kremlyovskaya 18, 420008 Kazan, Russia \\
}

\date{Accepted 2008 July 01. Received 2008 May 15; in original form 2008 February 11}

\pagerange{\pageref{firstpage}--\pageref{lastpage}}
\pubyear{2008}

\begin{document}

\maketitle

\label{firstpage}

\begin{abstract}
We discover a pronounced dependence of the strength of the soft X-ray orbital modulation and the spectral hardness in Cyg X-1 in the hard state on its superorbital phase. We find our results can be well modelled as a combination of two effects: the precession of the accretion disc (which appears to cause the superorbital flux modulation) and the orbital-phase dependent X-ray absorption in an accretion bulge, located at the accretion disc edge close  to the supergiant companion but displaced from the line connecting the stars by about $25\degr$. 
Our findings are supported by the distribution of the X-ray dips showing concentration towards zero superorbital phase, which corresponds to the bulge passing through the line of sight. We Fourier analyse our model, and find it explains the previous finding of asymmetric beat (between the orbital and superorbital modulations) frequencies in the observed power spectrum, provided the disc precession is prograde. On the other hand, we find no statistically significant changes of the orbital modulation with the superorbital phase in the 15-GHz radio data. This absence is consistent with the radio being emitted by a jet in the system, in which case the orbital modulation is caused by wind absorption far away from the disc. We also find that both the X-ray and radio fluxes of Cyg X-1 in the hard state on time scales $\ga\! 10^4$-s have lognormal distributions, which complements a previous finding of a lognormal flux distribution in the hard state on $\sim$1-s time scales. We point out that the lognormal character of the flux distribution requires that flux logarithms rather than fluxes themselves should be used for averaging and error analysis. We also provide a correct formula for the uncertainty of rms of a light curve for the case when the uncertainty is higher than the measurement. 
\end{abstract}

\section{Introduction}
\label{intro}

A number of X-ray binaries show flux periodicities at their respective orbital period, which may be caused by a number of effects. First, the source associated with the compact object in a binary may be eclipsed by the companion (usually of high mass) \citep[see e.g. a list ][]{wen06}. Second, a flux modulation may be caused by an optically-thick disc rim (which is highest at the point of impact of the gas stream from the inner Langrangial point in case of a donor filling its Roche lobe), obscuring the disc and/or its corona (e.g., \citealt{ws82,hm89}). This obscuration may lead to strong partial eclipses in so-called X-ray dippers. More generally, the disc and any associated structures may depart from its axial symmetry due to the influence of the companion, which may cause an orbital modulation. Third, wind from a high-mass companion may absorb/scatter the emission from the vicinity of the compact object, and the degree of absorption will depend on the orbital phase. In the case of Cyg X-1, both X-ray and radio emission are modulated by this effect, which modulations were modelled by, e.g., \citet{wen99} and \citet{sz07}, respectively. Fourth, phase-dependent absorption (via photon-photon pair production) of high-energy \g-rays may occur in a photon field axially asymmetric with respect to the compact object, especially that of the stellar photons (e.g., \citealt{bednarek06}). A fifth effect of the companion is reflection or reprocessing of the emission from around the compact object on the surface of the companion facing the compact object. This effects appears to be responsible for, e.g., the UV flux modulation from the X-ray binary 4U 1820--303 \citep{ak93,anderson97}. Finally, the optical/UV emission of the companion will be modulated if its shape departs from the spherical symmetry by partially or fully filling its Roche lobe, which effect is seen in Cyg X-1, e.g., \citet{brock2}.

Then, there will be an intrinsic dependence of the emitted flux on the orbital phase if the orbit is elliptical. This leads, e.g., to periodic outbursts at the periastron of Cir X-1 \citep{parkinson03} and Be/X-ray binaries  (see, e.g., \citealt{coe00,negueruela04} for reviews) in X-rays, and sometimes, at other wavelengths.  Also, some orbital flux modulation may be due to the Doppler effect, which is in principle observable \citep{ps87}, but has not yet been detected in a binary. (Obviously, the Doppler effect leads to widely observed shifts of spectral lines from binaries.)

In addition, a number of X-ray binaries show modulation at periods much longer than their orbital periods, so-called superorbital periodicity, see, e.g., a partial list in \citet{wen06}. In particular, Cyg X-1 shows such periodicity with the period of $\sim$150 d (e.g., \citealt{brock,kar01,od01,l06}, hereafter L06; \citealt{i07}, hereafter Paper I). The observed superorbital variability appears in most cases compatible with being caused by accretion disc and/or jet precession, which either results in variable obscuration of emitted X-rays as in Her X-1 \citep{k73}, or changes the viewing angle of the presumed anisotropic emitter, as in SS 433 \citep{k80} or Cyg X-1 (e.g., L06, Paper I), or both. The only known exception, in which the superorbital periodicity is clearly caused by modulation of the accretion rate (and thus not by a changing viewing angle of the source), is 4U 1820--303 \citep*{z07a}. 

A number of binaries show both orbital and superorbital modulations. Those currently known are LMC X-4, 2S 0114+650, SMC X-1, Her X-1, SS 433, 4U 1820--303 and Cyg X-1. An interesting issue then is whether there is any dependence of the parameters of the orbital modulation on the superorbital phase (or, similarly, on an average of the flux level). The shape of the profile of the orbital modulation in Her X-1 was found to depend on its superorbital phase \citep{sl99}, which appears to be due to the shadowing effect of the precessing accretion disc and scattering in its wind in that system. Recently, analogous dependencies of the shape of the orbital modulation on the average flux level have been found in LMC X-4, SMC X-1, Her X-1, as well as in Cen X-3 \citep{rp08}. Then, \citet{z07b} found such a dependence in 4U 1820--303 (of both the amplitude and the phase of the minimum flux) and interpreted it in terms of the size of the disc rim (partially obscuring the central source) changing with the variable accretion rate. 

In addition, there is the case of the peculiar Be/X-ray binary LS I +61$\degr$303, which shows orbital variability in the radio, X-ray and TeV emission, and a superorbital variability of the peak radio flux during an orbit \citep*{gregory99,gregory02}. \citet{gregory02} found a marked dependence of the phase of the peak of the orbital radio modulation on the superorbital phase in LS I +61$\degr$303. The presence of such a dependence may be due to interaction of the pulsar in that system with a variable circumstellar Be decretion disc \citep*{gregory02,z08}.

It is of considerable interest to find out whether orbital modulation depends on the superorbital phase in Cyg X-1, the archetypical and very well studied black-hole system with a high-mass companion, the OB supergiant HDE 226868 \citep{walborn}. In this work, we study this issue and find that such dependence exists and is very strong in soft X-rays. We then explain it theoretically in terms of orbital-phase dependent absorption in the stellar wind interacting with the outer accretion disc. 

\begin{figure}
\centerline{\epsfig{file= 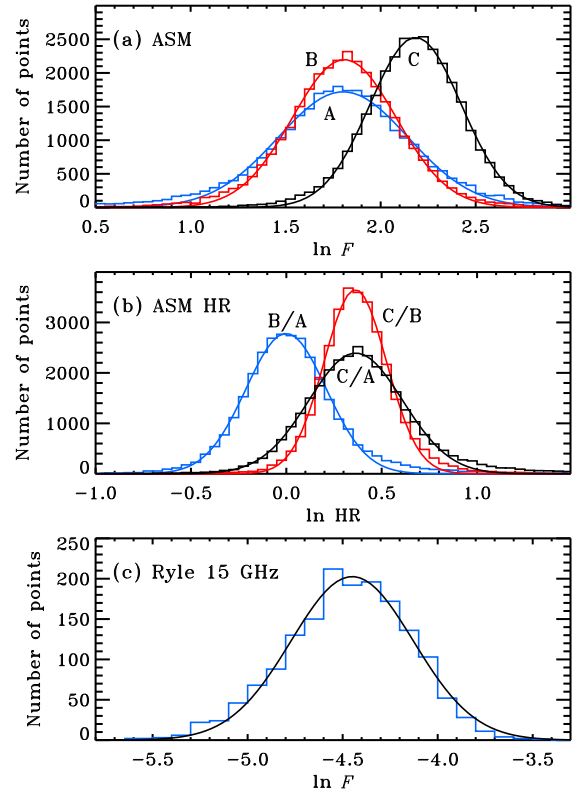,width=7.5cm}}
\caption{ 
(a) The histogram of the count rates (dwell-by-dwell data) observed from Cyg X-1 for the ASM A, B, and C detector channels. 
(b) Distribution of the hardness ratios B/A, C/B, and C/A.
(c) Distribution of the flux in the Ryle data. 
The flux units for the ASM and Ryle data are count s$^{-1}$ and Jy, respectively.
The solid curves give the best-fitting lognormal distributions.
}
\label{fig:histogram}
\end{figure}

\section{The light curves and their analysis}
\label{sec:method}

\subsection{Data}

We use the X-ray dwell data  (MJD 50087-53789, i.e., 1996 January 5--2006 February 23; note a misprint in the start date in Paper I) obtained with three Scanning Shadow Cameras of the All-Sky Monitor (ASM) aboard {\it Rossi X-ray Timing Explorer\/} (\xte; \citealt*{brs93,lev96}), with the channels A, B, and C corresponding to the photon energy intervals of 1.5--3 keV, 3--5 keV, and 5--12 keV, respectively. We also use the corresponding 15-GHz radio data from the Ryle Telescope of the 
Mullard Radio Astronomy Observatory (see, e.g., \citealt*{poo99}; L06 for earlier analyses of the observations of Cyg X-1). 
 
Because Cyg X-1 is a highly variable source and the effects we search for are  rather weak, we need to select accurately a homogeneous set of data. For most of the analysis in the paper, we  use  the  data corresponding to the hard spectral state following the criteria defined in section 2 of Paper I. We require the average photon spectral index derived from the \xte/ASM fluxes to be $< 2.1$ \citep{z02}, and additionally we exclude hard-state intervals with high X-ray variability, namely we  include only those 30-d intervals of the ASM data where $< 40$ per cent of points exceed by $4\sigma$ the average flux in the reference interval MJD 50660--50990. This has resulted in considering the following time intervals: MJD 50350--50590, 50660--50995, 51025--51400,   51640--51840, 51960--52100, 52565--52770, 52880--52975, 53115--53174, 53554--53690 (see fig. 1 in Paper I).

\subsection{Mean fluxes and variance}

We generallly follow the method of analyzing light curves described in Paper I, but with some modifications necessitated by the scientific goal of the present work. We use the orbital ephemeris of \citet{brock2} and the superorbital ephemeris of L06, see equations (1) and (4), respectively, in Paper I. We use the values of the orbital and superorbital periods of $P=5.599829$ d and $\psup=151.43$ d, respectively. We first divide an analyzed light curve into bins with the length of $P/20$. Then we average all points falling into a given bin weighted by the inverse squares of their measurement errors, obtaining the binned light curve, $F_i$. In this way, we avoid any contribution to our folded/averaged light curves from the source variability on time scales shorter than that corresponding to the length of our chosen phase bin (see Paper I). Note that unlike the method in Paper I, we do not prewhiten the light curves, i.e., do not subtract variability at one period in order to detect more clearly variability at another period. 

\begin{figure*}
\centerline{\epsfig{file=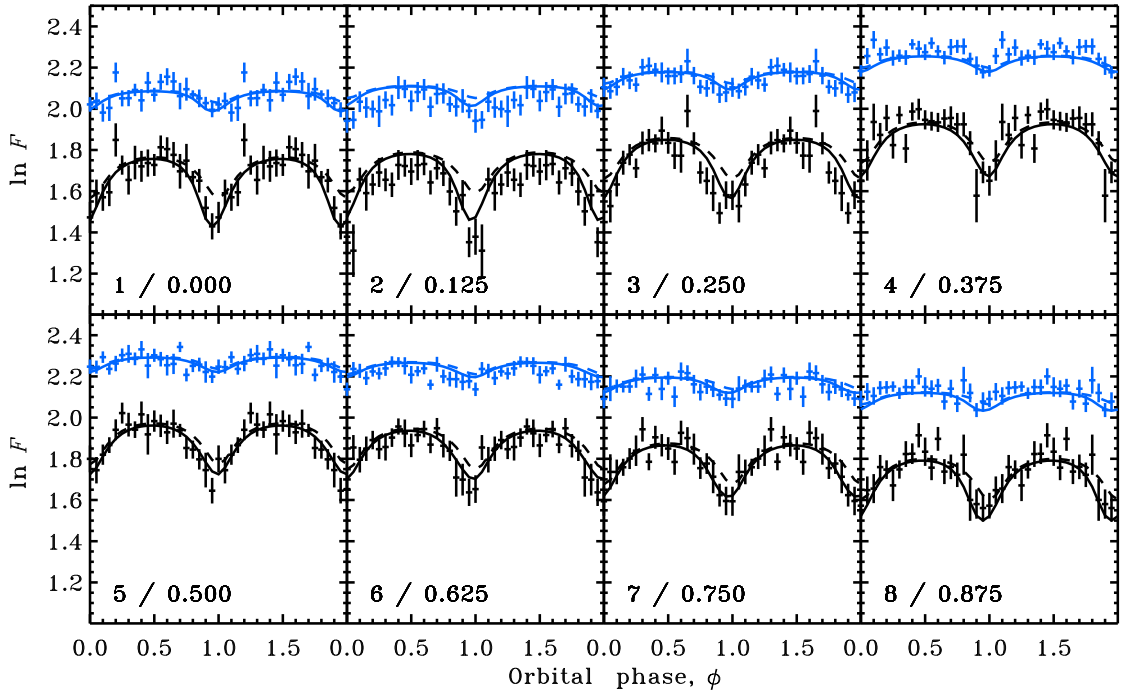,width=15.0cm}}
\caption{Profiles of the orbital modulation of the ASM A (1.5--3 keV, lower crosses) and C (5--12 keV, upper crosses) data at 8 superorbital phase bins. The bin number and the phase of the bin center are given on each panel. Note that the minimum and maximum modulation are offset from the 0 and 0.5 phase, and appear instead close to $\phisup=0.125$ and 0.625, respectively. The unit of $F$ is count s$^{-1}$. The solid curves give the best-fitting theoretical outflow model (model 8 in Table \ref{tab:fits}) described in Section \ref{sec:interpretation}, which involves the absorption in the isotropic stellar wind as well as in the bulge situated at the disc edge. The dashed curves shows the model component due to the wind only. 
}
\label{fig:profiles}
\end{figure*}

We have then looked into statistical properties of our distributions. We plot histograms of the fluxes for the ASM and Ryle data in Fig.\ \ref{fig:histogram}. We see that each of the histograms follows a lognormal distribution and it is completely inconsistent with a normal one. Our finding of the lognormal form of the variability of Cyg X-1 in the hard state on long time scales ($\sim$1/10-d to yr) in both X-rays and radio is supplemental to that of \citet*{uttley05}, who found the same type of distribution in X-rays on short time scales, $\sim$0.1--10 s, also in the hard state. This form of the flux distribution has important implications for calculating flux averages and the intrinsic dispersion, i.e., the standard deviation in the data. Namely, the standard-deviation error estimate based on the rms, namely 
\be 
\sigma^2= {\sum_{i=1}^N \left(x_i-\bar x \right)^2 \over N(N-1)},
\label{eq:sigma}
\ee
provides an unbiased estimate of the true standard-deviation error of the average of $x_i$ only if the distribution of $x_i$ is normal \citep{bevington92}. Therefore, for the purpose of calculating the averages and the rms standard deviations for our light curves, we have converted the count rates or fluxes in our binned light curves, $F_i$, into its logarithm, $G_i=\ln F_i$, with $G_i$ having now the distributions close to normal. 

We then separate the light curves binned based on the orbital phase into superobital phase bins of the length of $\psup/8$, with the mid-point of the first and the fifth bin at $\phisup=0$ and 0.5, respectively. Here, either the orbital phase, $\phiorb$, or superorbital phase, $\phisup$, is defined in the 0--1 interval, and 0 corresponds to the flux minimum as defined by the respective ephemeris. Then, we calculate folded and averaged profiles (of $G_i=\ln F_i$) of the orbital modulation within each superorbital phase bin, i.e.,
\be 
G_{jk}={\sum_{i\in(j,k)} G_i\over I_{jk}},
\label{G_jk}
\ee
where $i\in(j,k)$ counts over all points, $i$, falling into a given superorbital bin, $j$, {\it and\/} the orbital bin, $k$, and $I_{jk}$ is the number of such points. We estimate the error of this average using equation (\ref{eq:sigma}), i.e.,
\be 
\sigma_{jk}^2= {\sum_{i\in(j,k)} \left(G_i-G_{jk}\right)^2 \over I_{jk}(I_{jk}-1)},
\label{sigma_jk}
\ee
Note that this error estimate accounts for both the aperiodic variability of the source, i.e., intrinsic dispersion of individual fluxes contributing to a given orbital/superorbital bin (usually dominating), and the dispersion due to measurement errors. Also, since we use logarithms, $\sigma_{jk}$ represents a fractional error (and should not be divided by $G_{jk}$). 

The average and the average square error in a given superorbital bin are,
\be 
\bar G_j ={\sum_{k=1}^K G_{jk}\over K},\qquad
\bar \sigma^2_j ={\sum_{k=1}^K \sigma^2_{jk}\over K},
\label{averages}
\ee
respectively, where $K=20$ is the number of orbital bins.

We need to characterize the strength of a given modulation. One way of doing it without making any assumptions about its shape is to measure the fractional rms of a given orbital modulation profile. To do it, we calculate the unweighted rms variance and then subtract from it the rms variance due to the uncertainties of the individual points, which is so-called excess variance \citep[see e.g.][]{edelson02}),
\be 
S_j^2= {\sum_{k=1}^K \left(G_{jk}-\bar G_j \right)^2 \over K-1} -\bar \sigma^2_j .
\label{S2}
\ee 
Note that the variance difference above can be negative if the intrinsic variability is comparable or weaker than the measurement uncertainties. If this happens, we set this excess variance to zero. We again point out that $S_j$ represent already the fractional rms, i.e., it should not be further divided by $\bar G_j$ (which may be zero or negative). Then, we calculate the standard deviation of the above excess variance, $\Delta S^2_j$, following equation (11) of \citet{vaughan03}, hereafter V03,
\be 
\Delta S^2_j= \left(2\over K\right)^{1/2} \bar\sigma_j^2
\left(1+ {2 S_j^2\over \bar\sigma_j^2} \right)^{1/2}.
\label{Delta_S2}
\ee
We note that the transformation of $\Delta S^2_j$ into $\Delta S_j$ is not trivial. V03 have done it using the standard differential propagation of errors, obtaining their equations (B2) and (B3), which, however, we find not generally correct. Namely, the assumption behind using derivatives in propaging errors is that the uncertainty is much lower than the estimated quantity. This is often not the case for the excess variance, which can be null for either weak intrinsic variability or measurement errors comparable with that variability, see equation (\ref{S2}), whereas its uncertainty is always $>0$. Then, the error-propagation formula used by V03, $\Delta S_j=\Delta S_j^2/({\rm d}S_j^2/{\rm d}S_j)$ (using our notation), obviously fails, leading to infinite uncertainties. The cause for that is the failure of the assumption of $\Delta S_j^2\ll S_j^2$. To account for that, we calculate the uncertainty on the rms without that assumption, i.e., directly from the defition of the 1-$\sigma$ uncertainty range as $S_j^2\pm \Delta S_j^2$,
\be 
\Delta S_j = (S_j^2+\Delta S_j^2)^{1/2} - S_j.
\label{Delta_S}
\ee
Here we have chosen the upper error, which is larger than the lower one, and which is the only one possible for $S_j^2<\Delta S_j^2$. For $\Delta S_j^2\ll S_j^2$, this becomes the usual $\Delta S_j\simeq \Delta S_j^2/(2S_j)$ (as in V03), which equals $\Delta S_j\simeq\bar\sigma_j/K^{1/2}$. On the other hand, for $\Delta S_j^2\gg S_j^2$, the result is $\Delta S_j\simeq (\Delta S_j^2)^{1/2}$, which should be used to correct the upper part of equation (B3) in V03, and which equals $\Delta S_j\simeq (2/K)^{1/4}\bar \sigma_j$ (in our notation). Hereafter, we use equation (\ref{Delta_S}) to estimate the rms uncertainty.

Note that the above uncertainty estimates are due to the measurement errors only, and they do not account for the long-term, red-noise, variability of the source properties (V03). This is a correct procedure for our sample containing most of the currently available ASM data, for which we are interested in their actual properties, and are not hypothesizing about their behaviour over time scales $\gg 10$ yr. 

\subsection{Hardness ratio}
\label{sec:hr}

We also would like to analyse the spectral variability of Cyg X-1 with orbital and superorbital phase. A useful measure of the spectral shape is the hardness ratio (HR) of the fluxes in various channels, which can be computed in a number of ways. The obvious one is to use already available mean fluxes and construct their ratio.  This procedure, however, does not account for short-timescale spectral variability. The HR can also be computed for each observation (dwell) and then the mean can be obtained. 
However, we have already seen that fluxes follow the log-normal distribution, and therefore expect that their ratio could also be distributed in such a way.  Indeed, Fig.\ \ref{fig:histogram}(b) demonstrates that the logarithm of HR have distributions close to normal. Therefore, for the unbiased estimation of the mean HR and its error, we take the logarithm of HR for each observation (using fluxes that are not pre-averaged within the $P/20$ bins)  and average them within selected orbital and superorbital phase bins. We also note that the mean HR is computed without weighting the individual HRs according to their errors, because the error is systematically larger for harder spectra (as a result of a smaller flux in lower-energy channels), and therefore accounting for errors would result in a strongly biased estimate of the mean.

\section{Strength of the orbital modulation vs.\ the superorbital phase}
\label{sec:dependence}

The folded and averaged profiles of the orbital modulation for the ASM A data are shown in Fig.\ \ref{fig:profiles}. We can see that the orbital modulation is variable, e.g., it appears to be the weakest at the superorbital phase $\phisup= 0.625$. However, there is also a fair amount of statistical noise, and the results of this figure need to be quantized. We can see here that the orbital modulation profiles are characterized by rather narrow minima, and thus would not be well fitted by a smooth function, e.g., a sinusoid. Thus, we first calculate the rms of each dependence to characterize its strength, following the method of Section \ref{sec:method}.

\begin{figure}
\centerline{\epsfig{file=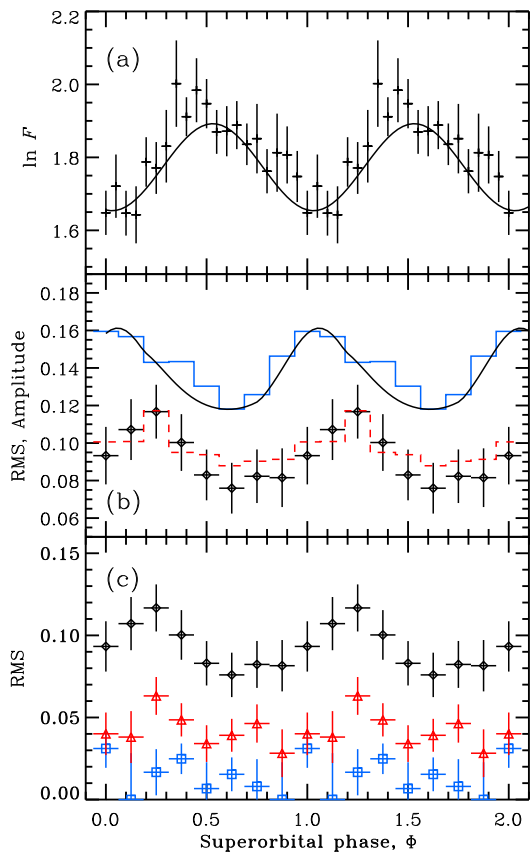,width=7.0cm}}
\caption{ 
(a) The superorbital phase diagram for the ASM A channel. The unit of $F$ is count s$^{-1}$. (b) Comparison of the characterization of the ASM A rms dependence using different methods.  The crosses are the intrinsic rms, $S$, of the orbital modulation as a function of the superorbital phase.  The solid histogram gives the amplitude of the orbital variability as fitted by sum of three harmonics, see Section \ref{sec:dependence}. The dashed histogram gives the corresponding rms for the fitting functions. The solid curves in panels (a) and (b) show the dependencies for the theoretical outflow model (model 5 in Table \ref{tab:fits}). (c) The dependencies of the intrinsic rms of the orbital modulation on the superorbital phase for three ASM channels. The crosses with filled circles, open triangles and open squares correspond to the channels A, B, and C, respectively. }
\label{fig:rms_asm}
\end{figure}

Fig.\ \ref{fig:rms_asm}(a) shows the superorbital phase diagram for the ASM A channel. We can see the highly significant flux modulation with the superorbital period (cf.\ L06, Paper I). We also see that the minimum of the superorbital cycle is clearly offset from the ephemeris of L06 by $\Delta\phisup\simeq 0.1$ (which was based on $\sim$30 yr of data compared to 10 yr analyzed by us). The crosses in Fig.\ \ref{fig:rms_asm}(b) show the corresponding rms dependence. We very clearly see a strong dependence of the rms on $\phisup$, with the rms being anticorrelated with the flux. It also appears that some phase lag, $\la 0.1$, of the maximum of the rms with respect to the minimum of the flux is present. 

Fig.\ \ref{fig:rms_asm}(c) shows the results for all three ASM channels. The orbital modulation, due to bound-free absorption, is strongest in the 1.5--3 keV range and weakest in the 5--12 keV range (\citealt{wen99}; L06). Consequently, the statistical significance of the dependence on $\phisup$ decreases with the energy.

\begin{figure}
\centerline{\epsfig{file=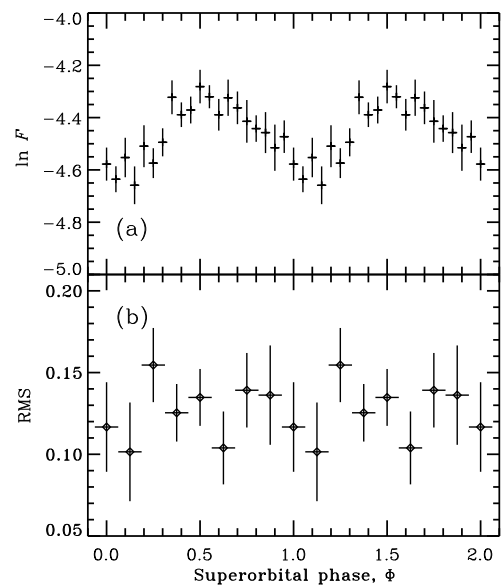,width=6.5cm}}
\caption{ 
(a) The superorbital phase diagram for the Ryle 15 GHz data. The unit of $F$ is Jy. 
(b) The dependence of the intrinsic rms of the 15 GHz orbital modulation on the superorbital phase, consistent with being constant.}
\label{fig:rms_ryle}
\end{figure}

In order to test the robustness of our finding of the dependence of the strength of the orbital modulation on $\phisup$, we have also calculated the rms for the ASM A taking into account the weights due to uncertainties of the individual points in the orbital phase diagrams (see \citealt{zdz04}). This alternative method gives only negligible differences with respect to the original one, and thus we do not show its results. Then, we have fitted the ASM A orbital modulation profiles with a sum of three sinusoidal harmonics, see equation (2) in Paper I, and calculated both the amplitude, $(F_\mathrm{max} -F_\mathrm{min})/(F_\mathrm{max}+ F_\mathrm{min})$, and the rms for it.  In this way, we largely avoid contributions to the rms from residual aperiodic variability. The results are shown in Fig.\ \ref{fig:rms_asm}(b). We see that the values of the rms of the fitted functions are very similar to that calculated directly from the data in Fig.\ \ref{fig:profiles}. On the other hand, the amplitude (which is sensitive only to the extremes of the fitted function) is larger than the rms simply due to their different definitions. The amplitude also shows a strong dependence of the superorbital phase similar in shape to that of the rms; however, it appears consistent with no phase shift with respect to the flux profile (Fig.\ \ref{fig:rms_asm}a). 

We have then searched for a similar effect in the Ryle 15 GHz data. We have found, however, that no apparent dependence is seen, and the $\phisup$-dependent orbital modulation profiles look all similar, and consistent with the average orbital modulation (see Fig.\ 4 in L06). Thus, we show here, in Fig.\ \ref{fig:rms_ryle}, only the results of calculating the rms of the orbital modulation as a function of $\phisup$. In Fig.\ \ref{fig:rms_ryle}(b), we see that the strength of the orbital modulation is consistent with being constant, though we cannot rule out some dependence hidden in the statistical noise. We have also checked that the 2.25 and 8.30 GHz data from the Green Bank Interferometer (see L06; Paper I) also do not show any statistically significant dependencies. 

\section{Spectral variability}
\label{sec:dips}

\subsection{Hardness ratio}

\begin{figure}
\centerline{\epsfig{file=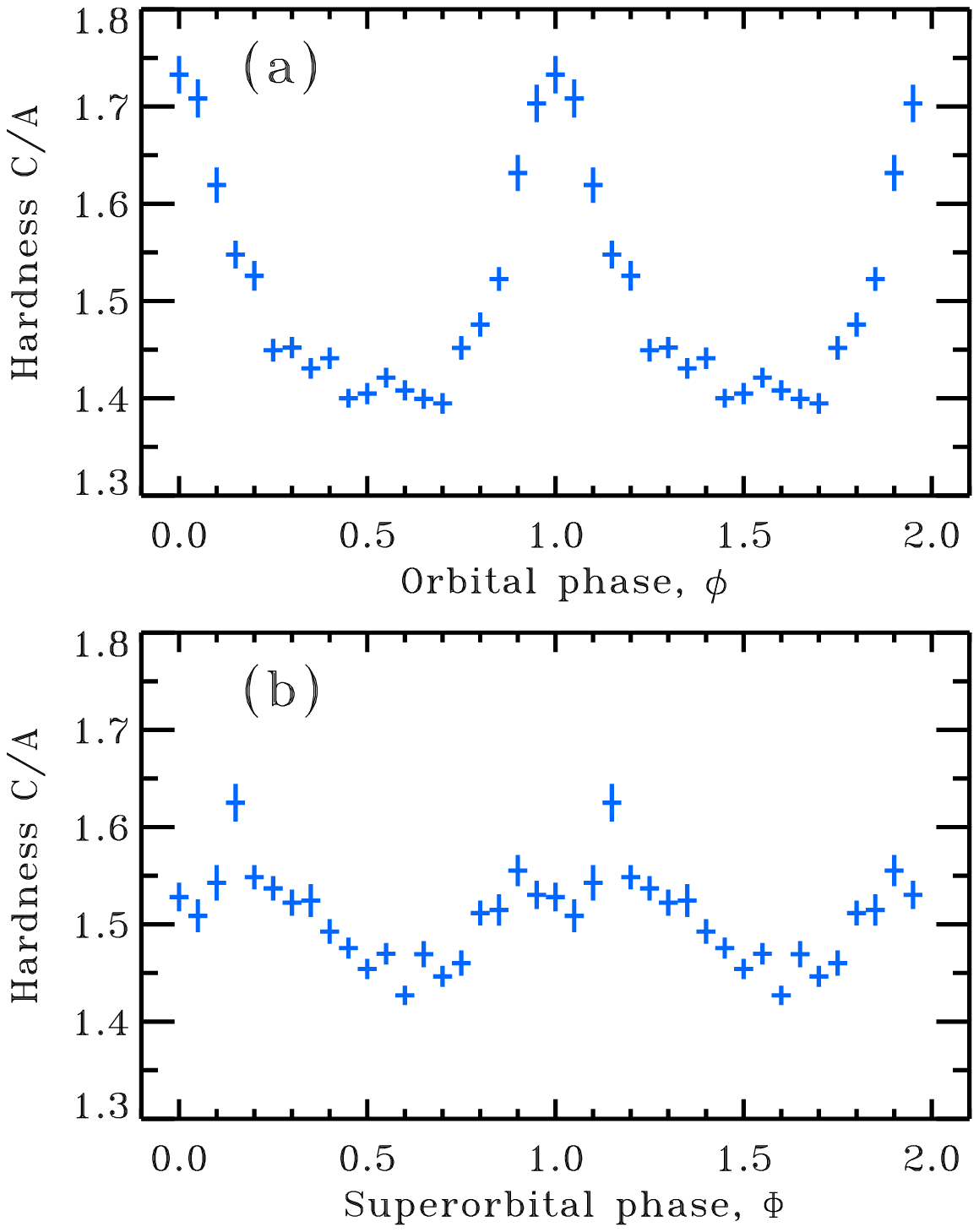,width=6.7cm}}
\vspace{0.5cm}
\centerline{\epsfig{file= 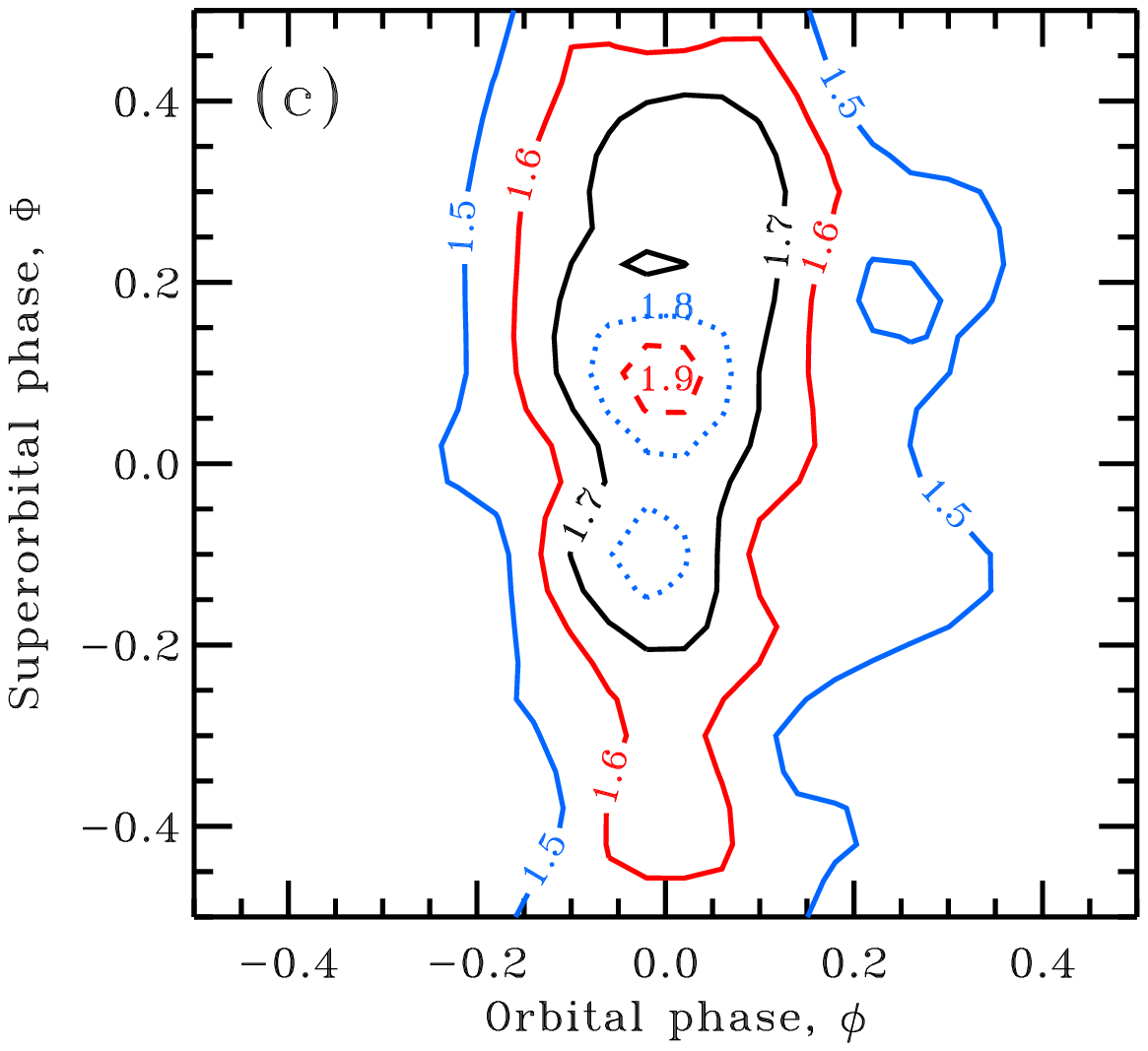,width=6.7cm}}
\caption{The mean hardness ratio in the ASM channels C and A in the hard state as a function of the (a) orbital and (b) superorbital phase. 
(c) Contour plot of the smoothed distribution  of the hardness ratio C/A over the orbital and superorbital phases. 
 }
\label{fig:hr}
\end{figure}

The X-ray modulations can also be tracked through the hardness ratio. The largest and easily detectable variability is shown by the ratio of count rates in ASM channels C and A (C/A). Fig.\ \ref{fig:hr} presents the dependence of the mean C/A (computed from the logarithm of the ratio, see Section \ref{sec:hr}) on orbital and superorbital phases.  We see a strong peak at orbital phase $\phiorb\sim 0$, which can be explained by absorption in the nearly isotropic wind \citep[see ][]{wen99}. The dependence of C/A on the superorbital phase also shows a very significant hardening around $\phisup\sim0$. The two-dimensional dependencies on $\phiorb$ and $\phisup$ demonstrate a plateau with C/A$\approx$1.4, a significant increase in hardness around $\phiorb=0.0\pm0.2$, and two peaks at superorbital phase $\phisup\sim-0.1$ and 0.1, which significance is not certain. Other hardness ratios C/B and B/A show similar behaviour, but of smaller amplitude.

\begin{figure}
\centerline{\epsfig{file=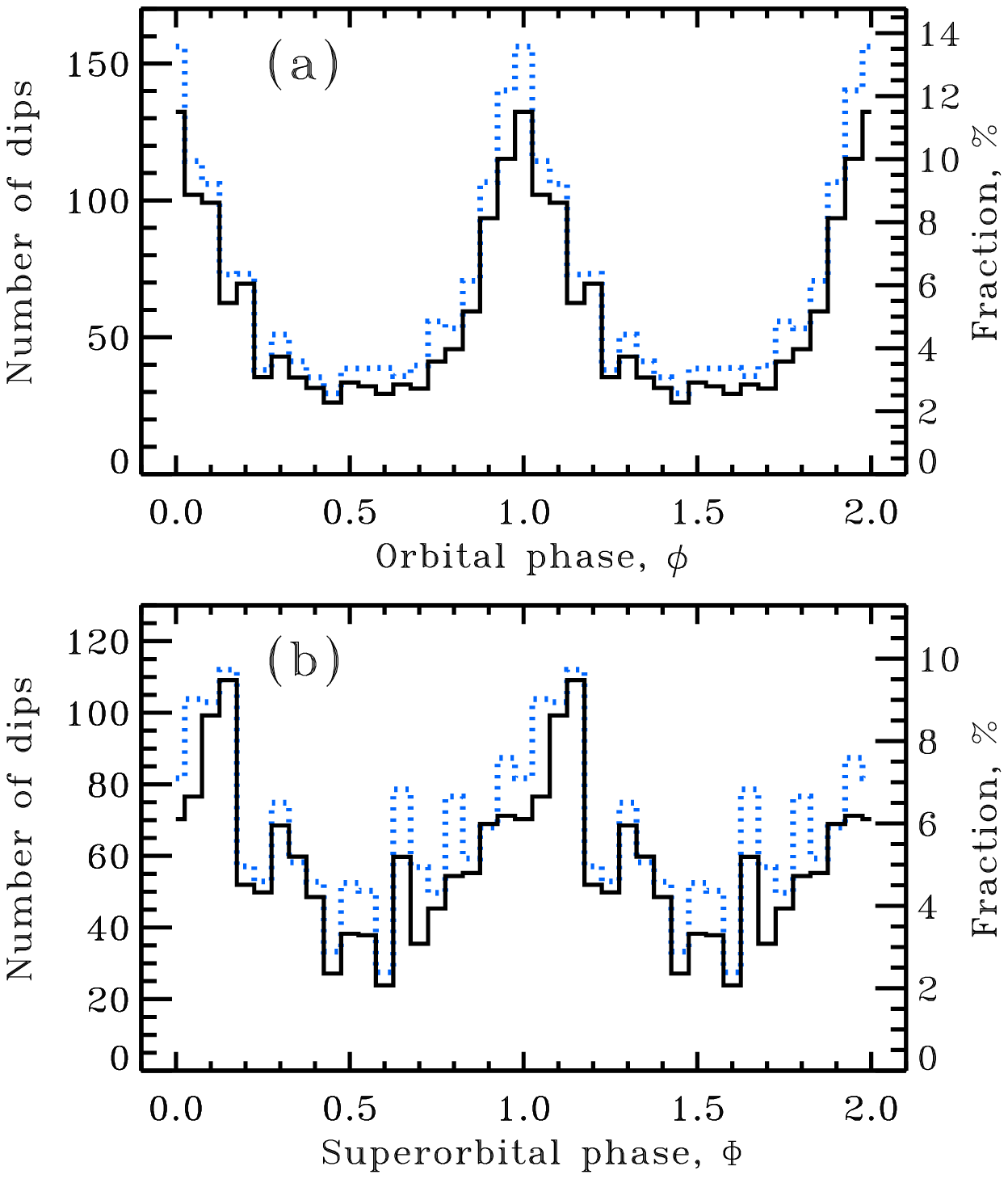,width=7.2cm}}
\vspace{0.5cm}
\centerline{\epsfig{file= 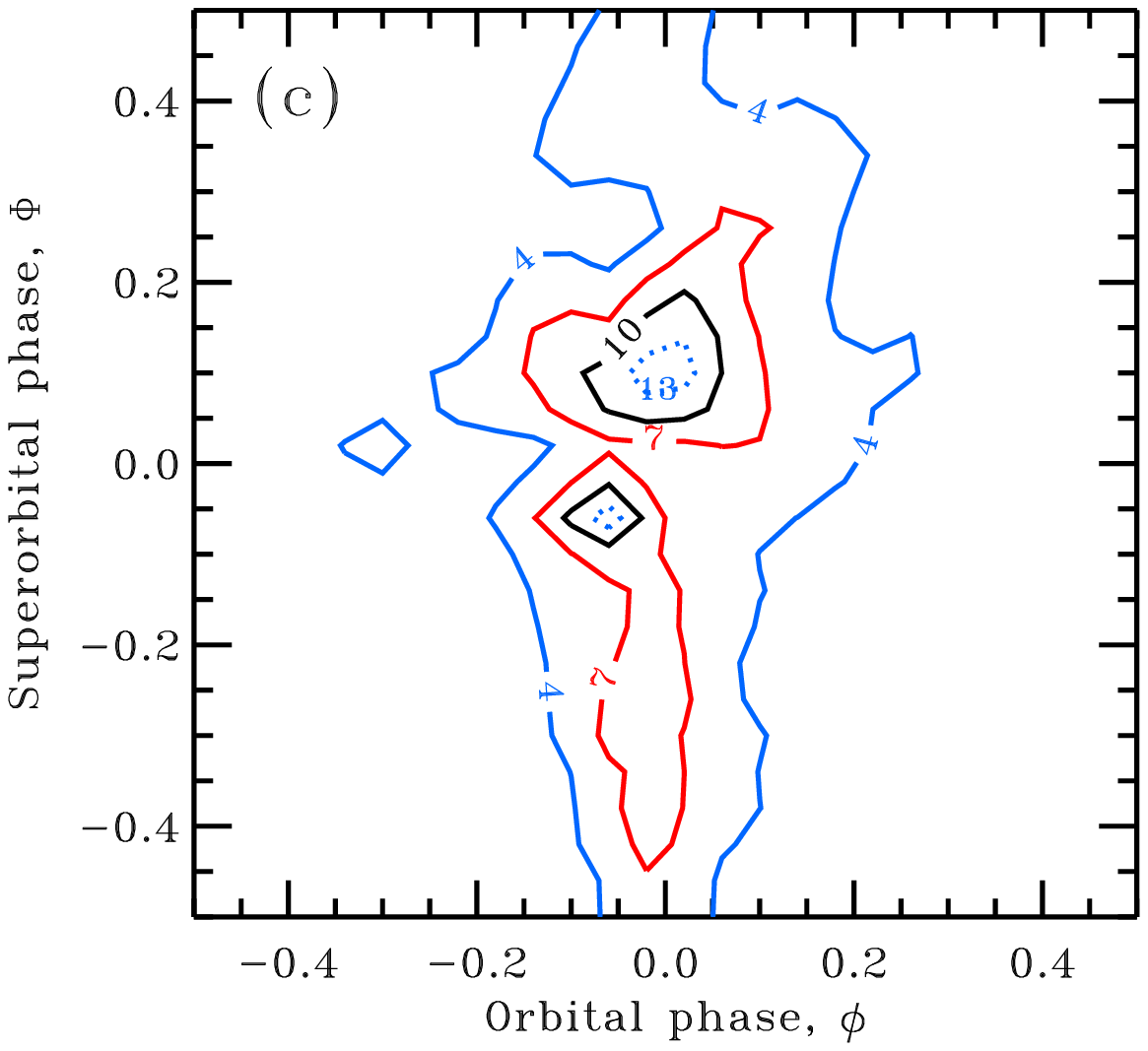,width=6.7cm}}
\caption{The distribution of X-ray dips over (a) orbital and (b) superorbital phase corrected for the coverage. The solid histogram is the hard state studied by us, while the  dashed histogram is for the entire ASM data set. The fraction scale corresponds to the solid histograms. (c) Contour plot of the smoothed distribution of all X-ray dips over the orbital and superorbital phases. }
\label{fig:dips}
\end{figure}

\subsection{X-ray dips}

X-ray dips, which are believed to result from absorption in blobs in the stellar wind, are characterised by significant drop in the count rate (see e.g. \citealt{bal00}, hereafter BC00; \citealt{fc02}). However, most markedly  they manifest themselves by spectral hardening (BC00). It is of interest to study their distribution over the orbital and superorbital phase and to compare these distribution to the corresponding dependencies of the HR. 

In order to define the dips, we use the ratio of the ASM count rates in channels B and A, B/A (HR1 in BC00), and the analogous C to B ratio, C/B (HR2 in BC00).  We then use the criteria of B/A$>$2 or C/B$>$2.5, which is similar to that of BC00 except that  they stated that they used {\it both\/} criteria simultaneously. With the present ASM calibration, we found only 56 dips satisfying their criterion in the ASM data used by us. The cause for almost no dips with both hardnesses large appears to be caused by the dip absorption being partial, i.e., with some small fraction of the flux remaining unabsorbed. Then, at a relatively low absorbing column, the flux in the A channel is reduced but the B and C channels are only weakly affected, so this case yields a B/A ratio increase but not C/B. On the other hand, at a column yielding a substantial reduction of the B flux, the C/B ratio increases, but the absorbed flux in the A channel is so low that it is dominated by the constant unabsorbed component, which results in no substantial increase of the B/A hardness. 

In the hard state data, we have found 1151 dips (814 with B/A$>$2 and 387 with C/B$>$2.5) among 31211 observations, while in the whole 10-year data set without any selection we have found 1336 X-ray dips (995 with B/A$>$2 and 437 with C/B$>$2.5)  among 60127 independent observations. 
Thus most of the dips happen during the hard state. This is expected because the spectral softening and increase of the luminosity in the soft state strongly increases the ionization level of the wind, which results in a weaker photoelectric absorption (BC00; \citealt{wen99}). This also strongly confirms the accuracy of our criterion defining the hard state. 

Fig.\ \ref{fig:dips}(a) shows the distribution of the dips over the orbital phase renormalized to the number of ASM observations in each bin. The picture looks relatively similar to fig.\ 5 in BC00 (based on $\sim$2 yr of the data, i.e., five times less than in our data set). The peak is at $\phiorb\simeq 0$, and it is relatively symmetric, especially for the hard-state data only. We have also checked that the distributions of the dips selected separately in the B/A and C/B look very similar. On the other hand, the additional peak at $\phiorb\simeq 0.6$ claimed by BC00 is not found by us, and appears to be due to a statistical fluctuation in the previous data set. Indeed, the total number of counts in the three bins forming that excess was 34, whereas the continuum level (i.e, without the excess) in those three bins corresponds to about 25. Thus, the excess corresponds to only $\sim\! 1.5\sigma$ in the Poisson statistics. 
The existence of the feature at $\phiorb\simeq 0.6$ is also not supported by the dependence of the HR, which shows no signs of spectral hardening at this phase (see Fig.\ \ref{fig:hr}a,c). 

Then we have studied the distribution of the X-ray dips over the superorbital phase. The results  are shown in Fig.\ \ref{fig:dips}(b). We see a maximum around $\phisup\simeq 0.05$--0.1, which is consistent with the position of the flux minimum (see Fig.\ \ref{fig:rms_asm}a). The distribution is clearly asymmetric relative to the peak, with a slower rise and faster decline, and it looks like the inverted flux (i.e., $-\ln F$) of Fig.\ \ref{fig:rms_asm}(a). 

Then, the two-dimensional   distribution of the dips in $\phiorb$ and $\phisup$ is shown in Fig.\ \ref{fig:dips}(c). We see that most of the dips that give rise to the peak in the orbital phase distribution around $\phiorb\simeq 0.0\pm 0.2$ happen around the superorbital phase of $\phisup\simeq 0.1\pm0.2$. (The statistical significance of the presence of two, rather than one, separate peaks there is rather low, $\sim\! 2\sigma$.) The distributions of the dips resemble strongly that of the HR, which is natural because the dips just represent a tail of the HR distribution.

\section{Theoretical interpretation}
\label{sec:interpretation}

\subsection{Wind geometry in Cyg X-1}
\label{geometry}

\begin{figure}
\centerline{\epsfig{file= 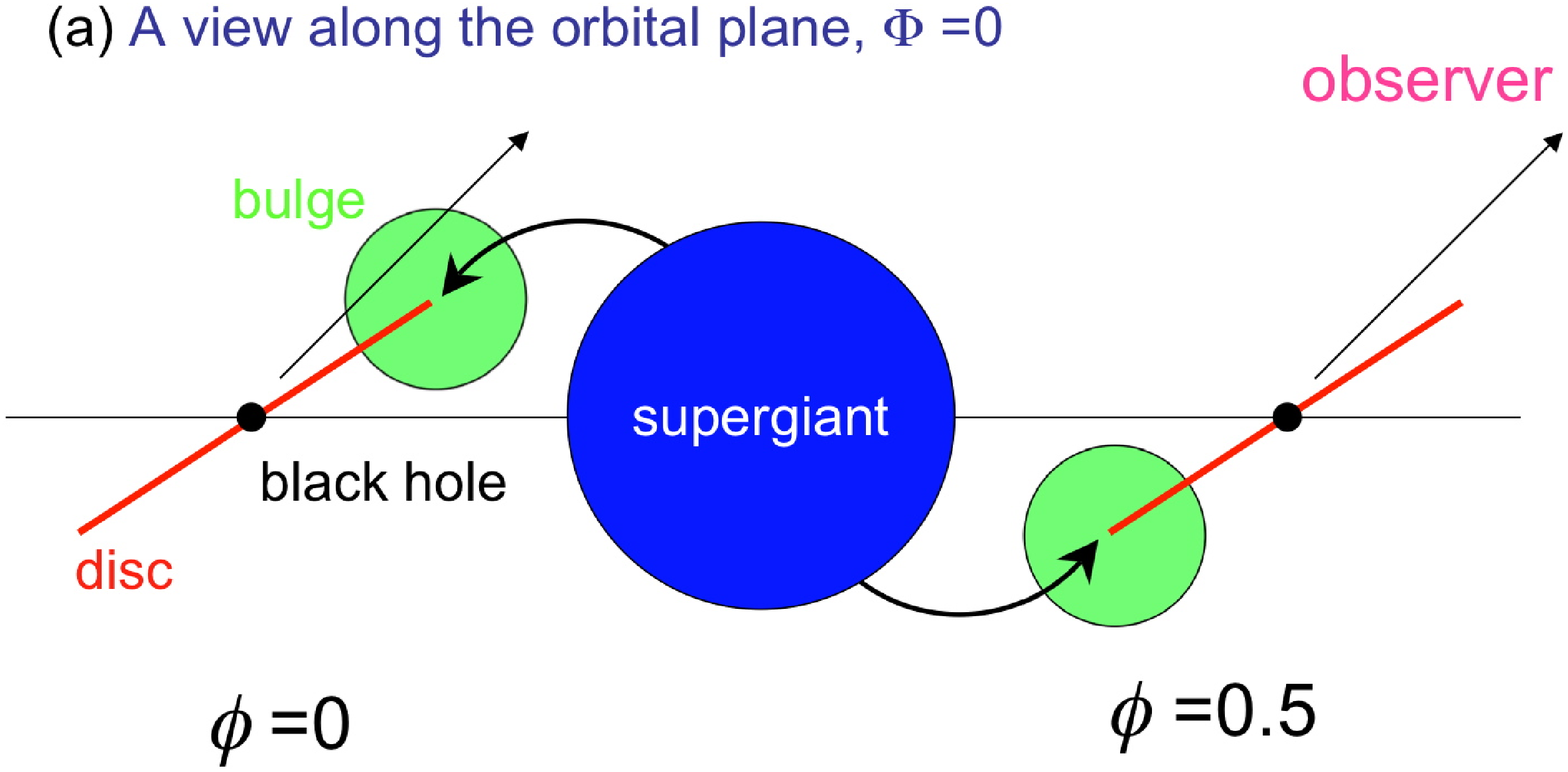,width=7.5cm}}
\vspace{0.2cm}
\centerline{\epsfig{file= 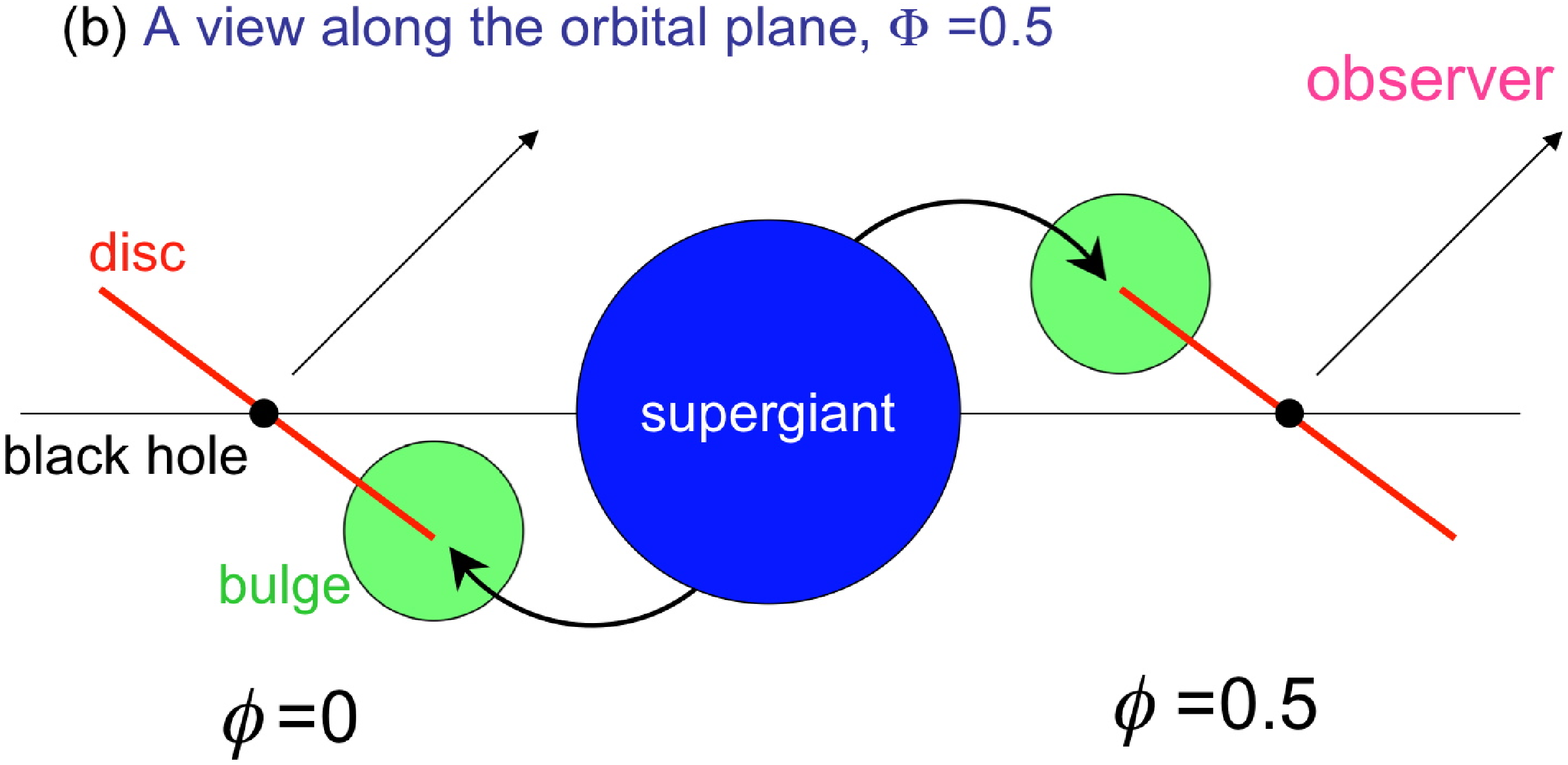,width=7.5cm}}
\vspace{0.2cm}
\centerline{\epsfig{file= 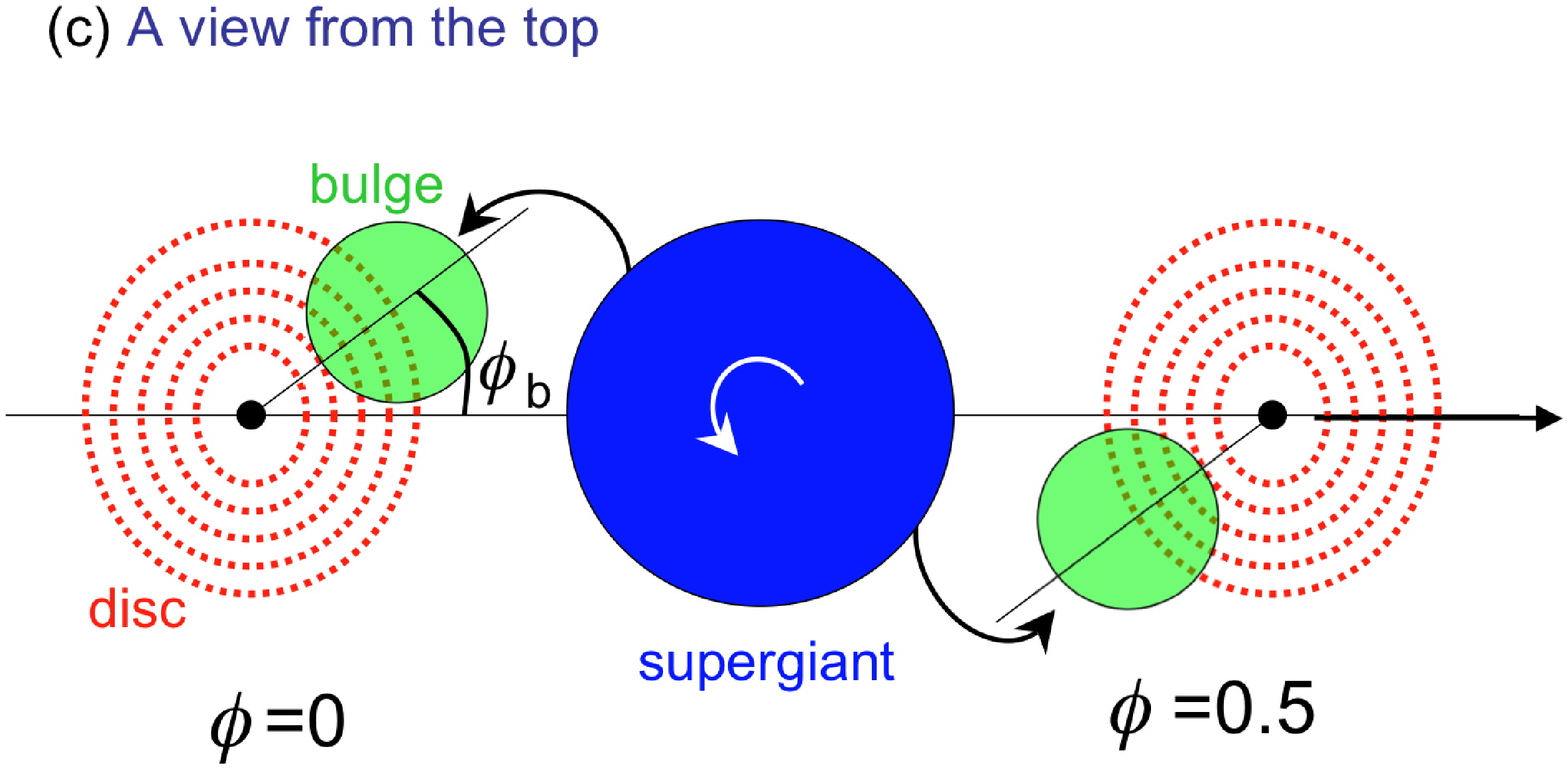,width=7.5cm}}
\caption{A drawing illustrating the effect of a bulge at the outer edge of a precessing inclined disc. The material in the bulge absorbs some of the X-ray emission originating close to the disc center. The orbital modulation due to the bulge is seen to strongly depend on the superorbital phase. In addition, there will also be orbital modulation due to the direct wind from the supergiant, not shown here for clarity. The elongation of the supergiant, almost filling its Roche lobe, is not shown here. A view along the orbital plane: (a) the superorbital phase of 0, when the disc is seen closest to edge-on and the effect of the bulge is strongest; (b) the opposite case of the superorbital phase of 0.5. (c) A view from the top, with the arrow showing the direction of the observer. The angle $\orbshift$ gives the azimuthal displacement of the bulge centre relative to the line connecting the stars, and it is $>0$ in the case shown here. The maximum of the absorption corresponds then to $\phiorb=-\orbshift$. This view is for any value of $\phisup$ except for the shown orientation of the elliptical image of the disc, which corresponds to $\phisup=0$ or 0.5. 
}
\label{fig:model}
\end{figure}

Let us first summarise our findings. In the radio, we see no modulation of the orbital variability with the superorbital phase, while in the X-rays such a modulation is visible. In addition to the previously known spectral hardening at orbital phase $\phiorb\sim0$ (visible in the HR and distribution of the X-ray dips), we find a significant increase in the HR around superorbital phase $\phisup=0$. This effect can be tracked in the dependence of the HR as well as the distribution of the X-ray dips. 


Our interpretation of the observed dependencies is as follows. 
The absence of statistically significant superorbital dependence of the orbital modulation of the 15-GHz radio emission is consistent with the radio being emitted by a jet in the system, in which case the orbital modulation is caused by wind absorption far away from the disc \citep{sz07}. For the X-rays, the situation is more complicated. 

The X-ray orbital modulation is due to variable absorption by the wind of the X-rays emitted close to the disc centre. The absorption can be separated into two components. One is independent of the superorbital modulation, and is due to absorption in the part of the wind steady in the comoving frame, as usually assumed. The other component is due to the part of the flow feeding the outer edge of the disc, and thus forming a bulge.

In Cyg X-1 system, though the OB star does not fill completely the Roche lobe, the wind density is enhanced inside the Roche lobe, which is an analog of the Roche lobe overflow but by the wind. Such a focused wind \citep{fc82,gies86b} in some way forms the accretion disc, known to exist in the system. The main argument for the existence of the disc is an overall similarity of the X-ray spectra and timing properties of Cyg X-1 to those of low-mass X-ray binaries, in which case accretion has to form a disc (see, e.g., \citealt{zg04}). The disc formation, most likely, leads to a condensation of the wind matter near the disc outer edge on the side of the companion in the form of a bulge, similar to the disc bulge inferred to be present in low-mass X-ray binaries, e.g., \citet{wh82,ws82,pw88,hm89}, as illustrated in Fig.\ \ref{fig:model}. On the other hand, the bulge can also be formed (see, e.g., \citealt{bor01}) by a shock wave in the wind when it encounters the gravity of the companion, the disc, or a wind from the disc, which is also likely to be present. In any case, when the fast, $>1000$ km s$^{-1}$, wind is stopped, the density increases dramatically. A fraction of the focused wind might pass  the black hole and be visible as additional absorption at orbital phase $\phiorb\sim0.5$, however we find no evidence for that in the X-ray data. 

An issue in the above scenario is the position of the bulge relative to the line connecting the stars. Consider the accretion process
in the corotating frame of the binary. In the case of low-mass X-ray binaries, the accretion stream leaves the L1 point with a small velocity and, being deflected by the Coriolis force, hits the disc (with  the outer edge defined by the stream orbital angular momentum)
at an azimuthal angle  $\orbshift\sim 60\degr$, which is measured from the line connecting the stars with the origin at the compact object (see Fig. \ref{fig:model}c  and the entry for  $\phi_h - 180\degr$  in table 2 in \citealt{ls75}). For the mass-ratio in Cyg X-1, $q = M_{\rm BH}/M_{\rm C} = 0.36\pm0.05$ \citep{gies03}, the gas freely falling from L1 point would hit the disc at  $\orbshift\sim 70\degr$. However, these considerations neglect the radiative acceleration of the stream as well as the diffusive spreading of the accretion disc and therefore its potentially much larger size, with both effects significantly reducing  $\orbshift$.  

An additional complexity is brought by a possibility of the non-synchroneous rotation of the companion in high-mass systems. For example, a slower stellar rotation allows the wind to be launched with a non-zero angular momentum in the corotating  frame and leads to the increase of  $\orbshift$, while the opposite is true for the faster rotation. The rotation of the companion in Cyg X-1 is compatible with corotation \citep{gies86a}, and therefore probably does not affect much the gas kinematics.  Then, if we measure this angle, $\orbshift$, in units of the 0--1 orbital phase, absorption of the X-ray emission in the bulge will peak at the orbital phase of $\phiorb\simeq 1-\orbshift$. Indeed, the typical phase of major X-ray dips in low-mass X-ray binaries is $\simeq 0.8$--0.9 \citep{pw88}. Some other high-mass X-ray binaries show dips at $\phiorb \simeq 0.8$--0.9, also thought to be caused by the accretion stream passing through the line of sight (\citealt{bor01} and references therein). 

A crucial further complication in Cyg X-1 is that the disc is inclined with respect to the binary plane and thus precesses. The precession causes changes of the position of the bulge with respect to the line of sight. During a single binary revolution, the bulge moves up and down, while the inclination of the disc remains approximately constant (since $\psup\gg P$), see Figs.\ \ref{fig:model}(a, b). At $\phisup$ close to zero, we see the disc at the highest angle, i.e., most edge-on. The  displacement of the bulge centre $\orbshift$ relative to the line connecting the stars  (see discussion above and Fig.\ \ref{fig:model}c) will also cause a small shift of the superorbital phase at which the bulge absorption is maximal. On the other hand, we see the disc close to face-on at $\phisup=0.5$, see Fig.\ \ref{fig:model}(b), and then the bulge is always outside the line of sight to the X-ray source. Thus, that additional absorption component is absent. 

The above considerations explain the dependence of hardness ratio on  orbital and superorbital phases as well as the distribution of the X-ray dips. Based on the two-dimensional distribution of the dips (Fig.\ \ref{fig:dips}c), we have calculated that at least 1/3 of all the X-ray dips are caused by the bulge, and the rest are due to the isotropic part of the stellar wind. The picture in Fig.\ \ref{fig:model} can also be used to calculate the expected X-ray orbital profiles caused by the wind and bulge absorption. We can assume a specific density profile of the wind and the bulge, and calculate the optical depth during a revolution for a given superorbital phase. 

\subsection{Model}
\label{sec:model}


Let us consider first the isotropic component of the wind. 
The wind mass density as a function of distance from the center of the star, $r$, can be estimated from the mass conservation law 
\be 
\rho_{\rm iso}(r) =\frac{\dot{M}}{4\pi r^2 v(r)},
\label{eq:rho_w} 
\ee
where $\dot M$ is the mass loss rate. We assume $v(r)\propto (1-R_*/r)^{\zeta}$, where $R_*$ is the stellar radius, 
and consider the attenuation cross-section independent of the distance. We thus get the absorption coefficient in the form 
\be
 \label{eq:abswind}
\alpha_{\rm iso}(r) = \alpha_{0} 
\left( \frac{a}{r} \right)^2 \left( \frac{1-R_*/a}{1-R_*/r} \right)^{\zeta}, 
\label{eq:alpha_w}
\ee
where $a$ is the separation between the black hole and the companion, and $\alpha_{0}$ is the absorption coefficient at $r=a$. 
We define here the characteristic optical depth, $\tauw =a\alpha_{0}$. 

\begin{figure}
\centerline{\epsfig{file=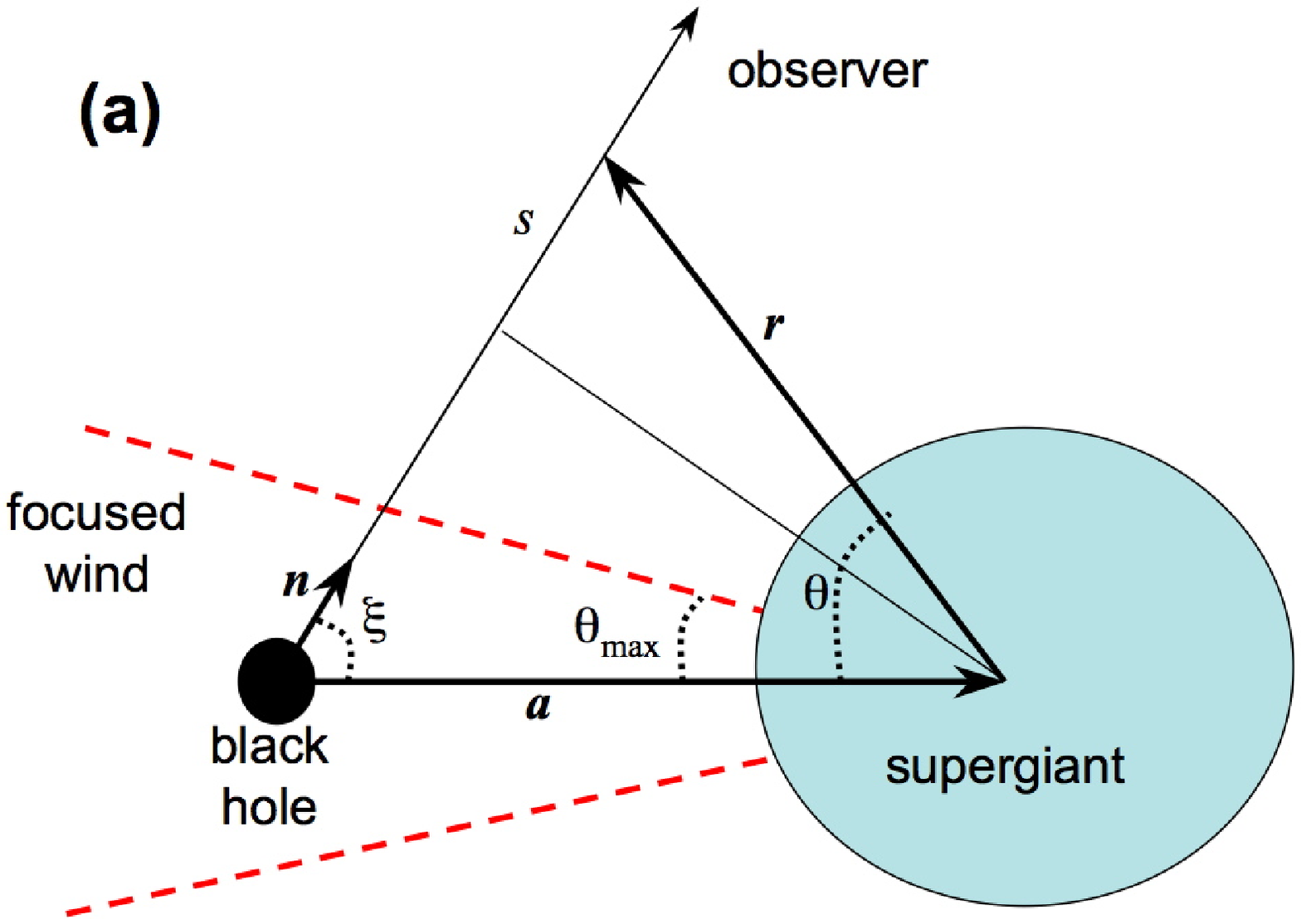,width=8cm}}
\vspace{0.2cm}
\centerline{\epsfig{file= 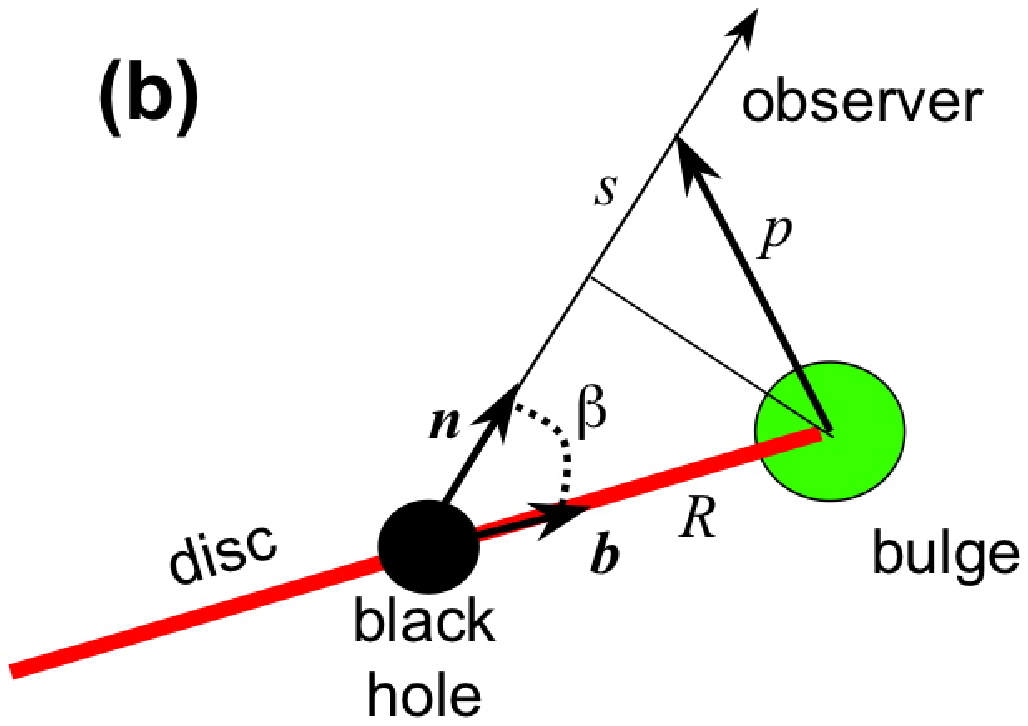,width=6cm}}
\caption{ (a) Geometry of the wind. (b) Geometry of the bulge. 
}
\label{fig:geom}
\end{figure}

The focused wind can be described by the cone of half-opening angle $\thetamax$ centred around the line connecting the stars (see Fig. \ref{fig:geom}a for geometry).  The additional opacity can be scaled to the opacity of the isotropic component and its angular 
dependence can be approximated   by a parabola \citep{fc82,gies86b}
\be \label{eq:focwind}
\alpha_{\rm fw}(r,\theta) = \alpha_{0} ( \etafw -1 ) \left[ 1 - \left( \frac{\theta}{\thetamax}  \right)^2   \right], \quad \theta<\thetamax, 
\label{eq:alpha_fw}
\ee
where  $\theta$ is the angle measured from the line connecting the stars and $\etafw$ is the ratio of the wind  density in the direction of the black hole to that of the isotropic component. The total wind absorption coefficient is defined by the sum  $\alpha_{\rm w}(r, \theta) =  \alpha_{\rm iso}(r)  + \alpha_{\rm fw}(r,\theta)$. 

Let us now compute the optical depth through the wind along the line of sight. It depends on the position of the observer. We introduce the coordinate system centred at the black hole with the $z$-axis along the normal to the orbital plane, and the observer in the $x$--$z$ plane, so that the direction to the observer is $\bmath{n} = (\sin i, 0, \cos i)$. Position of the companion is then $\bmath{a}=a(\cos \phiorb, \sin\phiorb, 0)$, where $\phiorb$ is the orbital phase.
The angle between the line of sight $\bmath{n}$ and $\bmath{a}$ varies with phase: 
\be 
\cos \xi = \bmath{n}\cdot \frac{\bmath{a}}{a}= \sin i \cos \phiorb . 
\label{eq:cos_xi}
\ee
The impact parameter is $a\sin\xi$ and the distance of some point  in the wind to the supergiant centre  is  $r=\sqrt{s^2+a^2 \sin^2 \xi}$, where $s$ is its distance to the point of the closest approach (which can be negative). The corresponding radius vector is $\bmath{r}= \bmath{n} (s + a \cos \xi) - \bmath{a}$ (see Fig. \ref{fig:geom}a). The angle between $\bmath{r}$ and $-\bmath{a}$ is then
\be 
\cos \theta = \frac{ \bmath{r}}{r} \cdot \frac{(-\bmath{a}) }{a} =  \frac{1}{r} \left( a\sin^2\xi - s \cos\xi\right) . 
\label{eq:cos_th}
\ee

The optical depth through the wind  is computed as 
\be 
\tau_{\rm w}(\phiorb) = \int_{-a\cos\xi}^{\infty}  \alpha_{\rm w}(r, \theta)  {\rm d}s . 
\label{eq:tau_w}
\ee

Let us apply this formalism to Cyg X-1. We take the ratio of the separation to the supergiant radius $a/R_* \approx 2.3$ \citep{zi05}, 
the inclination $i=40\degr$ (see Paper I and references therein),  the velocity profile exponent $\zeta=1.05$, and the focused wind parameters   $\thetamax=20\degr$ and  $\etafw=3$ \citep{fc82,gies86b}. In this case, the optical depth through the isotropic component of the wind from the black hole  to the infinity in the radial direction away from the companion is about $0.73\tauw$, in the perpendicular direction it is $1.26\tauw$ (i.e. at phases $\phiorb=0.25, 0.75$) and at zero orbital phase along the line of sight $\tau_{\rm w}(0)\approx 3 \tauw$. The typical optical depth provided additionally by the focused wind across its cone is $\approx \thetamax (\etafw -1)\frac{2}{3}  \tauw\approx 0.45 \tauw$. Thus  for Cyg X-1  at orbital phases $\phiorb\sim 0$, the focused wind adds only 15 per cent to the opacity produced by isotropic wind, while at $\phiorb\sim 0.5$, its contribution can reach 60 per cent, but the absorption itself at this phase is low. Thus in the first approximation, we can use only the isotropic wind model and include the corrections introduced by the focused wind later. 

In the case of the bulge, we first need to compute the position of the bulge centre, $\bmath{b}$, relative to the black hole. For the prograde precession  (see L06) the unit vector along the normal to the precessing accretion disc is $\bmath{d}= (-\sin\delta \cos\phisup, -\sin \delta\sin\phisup, \cos\delta)$, where $\delta$ is the precession angle. Assume now that the bulge centre lies  at the disc plane and the projection of  $\bmath{b}$   on the orbital plane $x$--$y$ makes  an  angle $\orbshift$ with the line connecting the black hole  to the companion (i.e. the azimuth of $\bmath{b}$ is $\phiorb+ \orbshift$, see Fig. \ref{fig:model}c).  We then get the unit vector of the bulge centre 
\begin{equation} 
\bmath{b} = \frac{[\cos(\phiorb+ \orbshift), \sin (\phiorb+ \orbshift), \tan\delta \cos(\phiorb+ \orbshift-\phisup)]}{\sqrt{1+\tan^2\delta \cos^2 (\phiorb+ \orbshift-\phisup)}}  . 
\label{eq:bulge_position}
\end{equation}
The angle it makes to the  line of sight is given by (see Fig. \ref{fig:geom}b)
\beq 
\cos\beta & =&  \bmath{b}\cdot \bmath{n} \\ 
& = &  \frac{\sin i \ \cos (\phiorb+ \orbshift) + 
\cos i \tan\delta\cos(\phiorb+ \orbshift-\phisup)   }
{\sqrt{1+\tan^2\delta \cos^2 (\phiorb+ \orbshift-\phisup) }}. \nonumber 
\label{eq:beta}
\eeq

Let us assume an exponential dependence of the absorption coefficient 
on the distance $p$ from the bulge centre, 
\begin{equation} \label{eq:absbulge}
\alpha_{\rm b}(p) = \alpha_{\rm b,0}  \exp(-p/\rb) ,
\label{eq:alpha_b}
\end{equation} 
with $\rb$ being the bulge scale-height.  This gives the  optical depth from the bulge centre to infinity of $\tau_{\rm b,0}=\rb\alpha_{\rm b,0}$. On the other hand, the optical depth from the black hole through the bulge along the line of sight,
\begin{equation} 
\tau_{\rm b}(\phiorb,\phisup) = \int_{-R\cos\beta}^{\infty}  \alpha_{\rm b}(p)  {\rm d} s , 
\end{equation}
depends on the orbital as well as superorbital phase. Here $R$ is the distance to the bulge centre from the black hole (i.e., approximately the disc size) and $p =\sqrt{s^2+ R^2 \sin^2 \beta}$.

For simplicity we assume that the wind and the bulge are independent and therefore the orbital modulation profile is given by 
\begin{equation} \label{eq:attenuation}
F(\phiorb, \phisup) = F_0(\cos\psi) \exp[-\tau_{\rm w}(\phiorb)]\  \exp[-\tau_{\rm b}(\phiorb,\phisup)] ,
\end{equation}
where $F_0$ is the intrinsic flux (which depends on $\phisup$) without absorption in the direction of the observer  and $\psi$ is the angle between the disc normal and the line of sight: 
\begin{equation}  \label{eq:cospsi}
\cos\psi= \bmath{n} \bmath{\cdot} \bmath{d}  = \cos i \cos \delta - \sin i  \sin \delta \cos\phisup .
\end{equation}
The retrograde precession can be modelled by substituting $\phisup\rightarrow-\phisup$ in the above formulae. 

\begin{sidewaystable}
\centering   
\rotcaption{Best-fitting model parameters.}
  \begin{tabular}{@{}crccccccrccl@{}}
  \hline
\# & Model$^a$ & $\delta$$^b$ & $\tauw$$^c$  & $\tau_{\rm b,0}$$^d$ & $\orbshift $$^e$ & $\supshift$$^f$ & $A$$^g$  & $\eta$, $\betaj$,  $\tau$$^h$ &  $\tau_{\rm C}/\tau_{\rm A}$$^i$ & $C/A$$^j$ 
& $\chi^2/{\rm dof}$$^k$ \\ 
 &  & deg                &   		     &   					   &  					       & 			   &                              &          &         \\
  \hline
1 & W+B a &  $7.5\pm0.5$    &  $0.09 \pm 0.03$ & $1.05^{+0.55}_{-0.44}$ & $0.07 \pm 0.04$  &  $0.03 \pm 0.02 $        & $9.2 \pm 0.2$ &  -- & &          &    $151.9/154$        \\
2 & W+B b &  10.0 (f)  		&  $0.09 \pm 0.02$ & $0.8^{+0.4}_{-0.35}$ & $0.08 \pm 0.03$  &  $0.04 \pm 0.01 $   & $11.6 \pm 0.8$ & $-0.27\pm0.06$ &   &          &  $155.2/154$          \\
3 & W+B c &   5.0 (f)        	&  $0.07 \pm 0.03$ & $1.35^{+0.65}_{-0.6}$	& $0.07 \pm 0.04 $  &  $0.03 \pm 0.01 $ 	& $2.90 \pm0.10 $ & $0.47\pm0.03$ &   &          &  $148.8/154$            \\
4 & W+B c &   7.5 (f)  	         &  $0.09 \pm0.03$ & $1.05^{+0.50}_{-0.45}$  & $0.07 \pm 0.04$  &  $0.03 \pm  0.01 $ 	& $3.55 \pm0.13 $ & $0.36\pm0.03$ &  &          &   $151.8/154$          \\
5 & W+B c &   10.0 (f)  		&  $0.09 \pm 0.02$ & $0.8^{+0.4}_{-0.35}$ & $0.08\pm 0.04$      &  $0.03 \pm 0.02 $      & $4.04 \pm 0.14$ & $0.29\pm0.02$ &  &          &   $154.7/154$         \\
6 &  W+B d   &  10.0 (f)  		&  $0.09 \pm 0.02$ & $0.8^{+0.4}_{-0.35}$ & $0.08 \pm 0.04$  &  $0.04 \pm 0.01 $ & $14.8\pm1.2$ & $0.56\pm0.05$ &   &          &  $155.9/154$         \\
7&  F+W+B c   &  10.0 (f)  		&  $0.09 \pm 0.02$ & $0.8^{+0.4}_{-0.35}$ & $0.08 \pm 0.04$  &  $0.03 \pm 0.01 $ & $4.20\pm0.15$ & $0.29\pm0.02$ &   &          &   $155.4/154$         \\
8 & W+B c &   10.0 (f)  		&  $0.09 \pm 0.02$ & $0.9^{+0.4}_{-0.35}$ & $0.08\pm 0.04$   &  $0.01 \pm 0.02 $ & $4.15\pm0.14$ & $0.27\pm0.02$ &   $0.3\pm0.1$  & $1.31\pm0.03$  & $359.2/318$        \\
\hline
\end{tabular}
\begin{flushleft}
{$^{a}$The models described in Section  \ref{sec:model}:  W is the isotropic wind model, F is the focused wind and B stands for the bulge. Small letters giving the models of the intrinsic emission from Section \ref{sec:fitting}. The model 8 is fitted to the  ASM A and C channels simultaneously.
$^{b}$The precession angle. 
$^{c}$Characteristic optical depth of the isotropic wind.
$^{d}$Characteristic optical depth of the bulge.
$^{e}$The shift of the bulge centre in orbital phase (fraction of the orbit).
$^{f}$The shift in superorbital phase.
$^{g}$The model normalization in the ASM A channel. 
$^{h}$The anisotropy parameter, the jet velocity, or the slab optical depth.
$^{i}$The ratio of absorption coefficients in channels C and A. 
$^{j}$The ratio of model normalizations in  channels C and A. 
$^{k}$$\chi^2$ and the number of degrees of freedom. The errors on the parameters  are given at 90 per cent confidence level for one parameter, i.e. for $\Delta\chi^2=2.71$. 
The size scale of the bulge in units of the disc size, $\rb/R$, is fixed at 0.2  and inclination $i$ is $40\degr$  in all of the models.}
\end{flushleft}
\label{tab:fits}
\end{sidewaystable}


\subsection{Modelling the data}
\label{sec:fitting}

In order to describe the profiles presented in Fig. \ref{fig:profiles} with the model of Section \ref{sec:model}, we need to specify the angular distribution of the intrinsic flux, $F_0(\cos\psi)$. In Paper I we have considered four simple analytical models: 

(a) the black body, with the flux proportional to the projected area, $F_0(\cos\psi)=A \cos\psi$. 
 
(b) an anisotropic model of $F_0(\cos\psi)=A \cos\psi (1+ \eta  \cos\psi)$ with parameter $\eta$ giving the degree of deviation from the black body. Such anisotropy can be produced for example by thermal Comptonization (Paper I; \citealt{st85,vp04}), which the dominant radiative process giving rise to X-rays in the hard state of Cyg X-1 \citep[e.g.,][]{gier97,pc98,p98}. 

(c) the steady jet model, $F_0(\cos\psi)=A [\gammaj(1- \betaj  \cos\psi)]^{-(1+\Gamma)}$, where $\betaj=v/c$ is the jet velocity, $\gammaj =1/\sqrt{1-\betaj^2}$ is the jet Lorentz factor, and $\Gamma$ is the photon index of the X-ray radiation. By the 'jet', we mean here  either the base of the jet in the direct vicinity of the black hole, or an outflowing corona \citep[see e.g.][]{b99,mbp01,mnw05}.  
 
(d) the slab absorption model, $F_0(\cos\psi)= A \exp(-\tau/\cos\psi)$, which can be associated, for example, with some kind of a disc outflow.

All the models provide a good fit to the superorbital variability of Cyg X-1 (Paper I). Models (b) and (c) can be considered as more physically motivated, but we consider here all of them. In order to keep the number of parameters to minimum we fix the inclination of the system $i=40\degr$. The precession angle $\delta$ is not well determined in models (b)--(d) as it is anticorrelated with other parameters ($\eta, \betaj, \tau$, see Paper I). Thus we fix it at three values  between $5\degr$ and $10\degr$. 

The parameters describing the absorption of radiation are the characteristic optical depths $\tauw$ and $\tau_{\rm b,0}$ for the wind and bulge, respectively. Additional parameters are  the bulge density scale measured in units of the disc size, $\rb/R$, and the phase shift, $\orbshift$, of the position of the bulge centre. An arbitrary shift in the superorbital phase, $\supshift$ (due to the uncertainty of the superorbital ephemeris), is also introduced (i.e. we replace  $\phisup$ by  $\phisup-\supshift$    in all formulae of Section \ref{sec:model}).  The parameters describing the radiation pattern are the normalization, $A$ (for ASM A channel), and the anisotropy parameter, $\eta$, in model (b), $\betaj$ in model (c) (where we fix $\Gamma$ at a typical hard-state value of 1.7), and the slab optical depth, $\tau$, in model (d).  

We consider the prograde precession of the disk (L06).
In order to understand the influence of the model complexity on the results, we consider first only the isotropic component of the wind (models W in Table \ref{tab:fits}) and fit the ASM A profiles only, which show strongest variability. 
We find that parameters $\rb/R$ and $\tau_{\rm b,0}$ are anticorrelated, and cannot be determined separately. This happens because various combinations of the two parameters can give the same optical depth  through the bulge at a given impact parameter. Therefore,  we fix $\rb/R=0.2$.  The best-fitting model parameters are presented in Table \ref{tab:fits}. For model (a), the precession angle agrees within the errors with the results of Paper I. The jet model, (c), provides a slightly better fit for smaller precession angles. The models (b) and (d) also give statistically similar fits. The phase shifts $\supshift$ and $\orbshift$ are well constrained by all the models. The fits require the shift of the  bulge centre  from the line connecting the stars by $\orbshift \approx0.07$ (i.e., $25 \degr$, see Fig.\ \ref{fig:model}c). All the models give similar optical depths through the wind and the bulge. The wind optical depth $\tau_{\rm w}$ varies between 0.28 (i.e. $\approx3\tauw$) and 0.08 (i.e. $\approx\tauw$)  for $\phiorb$ varying between 0  and 0.5. For the bulge, $\tau_{\rm b}$ varies between 0.15 and  0.007 at $\phisup=0$ and between 0.05  and 0.008 at $\phisup=0.5$. 

We now add an additional focused wind component with the parameters specified in Section \ref{sec:model} and fit the data using jet model (c). The resulting best-fitting parameters are not very much different  from those obtained with the isotropic wind model (compare entries 5 and 7 in Table 1). This is expected, because the focused wind  affects the total opacity on average at about a 30 per cent level. 

Finally, we fit the light curves in channels A and C simultaneously.  Two additional  parameters have to be introduced: the ratio of the absorption coefficients (and optical depths) in channels C and A, $\tau_{\rm C}/\tau_{\rm A}$, and the ratio  of the normalizations (intrinsic hardness ratio), $C/A$.  The best-fitting results for the main model parameters change only slightly (compare entries 5 and 8 in Table 1). Because the mean absorption coefficients in channels A and C differ  only by a factor of 3, the absorbing gas has to be rather strongly ionized. 

For the retrograde precession, all these models give much worse fits to the data.

\section{Discussion}

\subsection{The origin of beat frequencies}
\label{sec:beat}

\begin{figure}
\centerline{\epsfig{file=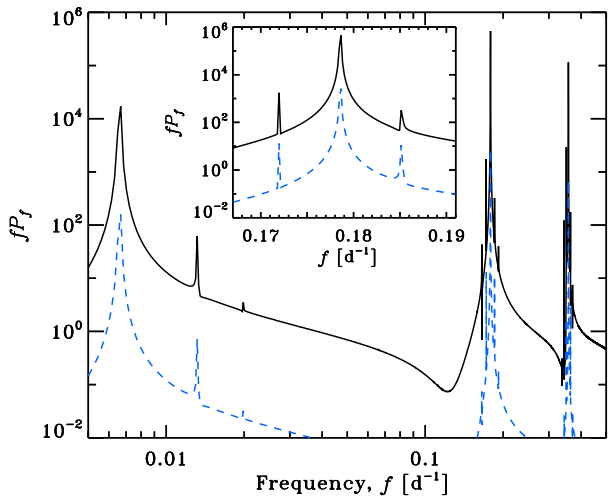,width=8cm}}
\caption{Power density spectra predicted by our models (arbitrary normalization). 
The solid curves show the power spectrum of the flux for the outflow model, 5 in Table \ref{tab:fits}. The inset zooms on the frequency range near $1/P$. The dashed curves show the model with the absorption only in the wind, i.e., neglecting the presence of the bulge.
}
\label{fig:fourier}
\end{figure}

A collateral effect of the coupling between the orbital and superorbital modulations may be appearance of additional frequencies in the power spectrum. If the two modulations were independent, there would be simply two peaks in the power spectrum at the corresponding frequencies. On the other hand, if one modulation depends on the other, beat frequencies, at $\nu=1/P\pm 1/\psup$, may appear. Indeed, L06 reported finding the lower of the beat frequencies  (albeit at a relatively limited statistical significance), and also found that its origin from X-ray reflection from the surface of the companion is unlikely. 

Here, we have tested whether the discovered dependence of the orbital modulation on the superorbital phase may indeed cause beat frequencies to appear. 
Using our model (given by equation (\ref{eq:attenuation}) and other formulae of Sections \ref{sec:model}, \ref{sec:fitting} with parameters of model  5 in Table \ref{tab:fits}), we have generated a light curve and computed Fourier power-density spectrum (PDS). We have found that our model gives rise to strong peaks at frequencies $1/\psup$ and $1/P$ with harmonics as well as to two peaks in the power spectrum at the beat frequencies. Interestingly, the lower beat-frequency peak is 3.7 to 5.4 times stronger than the higher one (depending on whether we compute Fourier transforms from the flux or from the logarithm of the flux). Fig.\ \ref{fig:fourier} shows the flux Fourier transform for this case for the outflow model. We then compare these predictions with a simpler model where absorption in the bulge is neglected. In this case, there are two beat-frequency peaks of equal strength in the PDS  of the flux (see the dashed curves in Fig.\ \ref{fig:fourier}), while they are missing in the PDS computed from the logarithm of the flux, because the coupling disappears.  If on the other hand, only bulge produces absorption, PDS shows both beat-frequency peaks with the strength ratio of 10 and 5 for the flux and its logarithm, respectively. Finally, we experimented with the model where intrinsic flux, $F_0$, as a function of superorbital phase was assumed to be constant and both bulge and wind are responsible for absorption. Now, the strength of the peak at $1/\psup$ has diminished by three orders of magnitude, while the behaviour of PDS at $1/P$ and the  beat frequencies was almost identical to the full model with variations of $F_0$ (the ratio of peak strengths is 6.5 and 5.5 for the flux and its logarithm, respectively). 

We see that the lower beat-frequency peak is always stronger than the higher one when absorption is modulated by the bulge (for prograde precession). The coupling discovered in this work thus predicts a presence of beat frequencies with a stronger low-frequency peak, which is consistent with the discovery by L06 of only the low-frequency peak. 

\subsection{Superorbital variability and outbursts of Cyg X-1}
\label{sec:outbursts}

It is of interest to consider whether the superorbital variability of Cyg X-1 is related to other aspects of the source activity. Recently, the MAGIC collaboration \citep{albert07} reported detecting TeV emission from Cyg X-1. That detection, on MJD 54002, took place in the middle of a strong X-ray outburst of Cyg X-1 \citep{t06}. We have checked that that time corresponds to the peak of the superorbital cycle, $\phisup\simeq 0.5$, when the disc and jet of Cyg X-1 are most face-on. On the other hand, L06 found that the superorbital cycle was uncorrelated with the appearance of other strong X-ray outbursts of the source of duration of days \citep{sbp01,gol03}. Thus, the significance of the coincidence of the TeV burst with the peak of the superorbital cycle in the present case remains unknown. 

Interestingly, the orbital phase of the TeV outburst was at $\phiorb\simeq 0.9$ \citep{albert07}, at which absorption of TeV photons by pair production on the stellar photons is very strong. A possible way to obtain detectable TeV emission is then via pair cascades. We note that the statistical significance of the detection was relatively limited, $4.1\sigma$, and thus an independent confirmation of this deteciton is desirable. 



\section{Conclusions}

We have discovered the dependence of the orbital modulation strength and the hardness ratio on the superorbital phase of Cyg X-1. The observed effects can be explained by the presence of the absorbing material more or less fixed in the corotating frame of the stars. We associate this material with the bulge formed by the accreting stream impacting the accretion disc. Because of the disc precession (causing superorbital variability), the bulge moves up and down and its influence on absorption varies. At the superorbital phase 0.5, the line of sight does not pass through the bulge, while at $\phisup\approx 0$ the absorption in the bulge is maximal. We estimate the maximal optical depth at 1.5--3 keV through the bulge (for our line of sight) of about 0.15, while the stellar wind produces twice as much of the absorption. 

Using a simple model of the bulge and the stellar wind incorporating the angular dependence of the intrinsic X-ray radiation from the black hole vicinity, we were able to reproduce the detailed shape of superorbital variability as well as of the orbital modulation at various superorbital phases. We find the bulge centre is displaced from the line connecting the stars by about $25\degr$. We also study the distribution of the X-ray dips over superobrital phase we find their concentration towards the superorbital phase $0.1$, which coincides  with the position of the flux minimum. We thus are in position to claim that the X-ray dips observed in Cyg X-1 at around zero orbital phase have direct relation to the bulge which, in turn, causes variation of the orbital modulation with the superorbital phase. We Fourier analyse our model, and find it explains the finding of only the lower beat frequency between the orbital and superorbital frequencies in the observed power spectrum (L06), provided the disc precession is prograde. 

We also find that both the X-ray and radio fluxes of Cyg X-1 in the hard state on time scales $\ga\! 10^4$-s have lognormal distributions, which complements the finding of a lognormal flux distribution in the hard state on $\sim$1-s time scales \citep{uttley05}. We stress out that the lognormal character of the flux distribution requires that flux logarithms rather than fluxes themselves should be used for averaging and error analysis. We also correct a mistake in the treatment of V03 of the uncertainty of intrinsic rms variability of light curves in the case when the uncertainty is higher than the intrinsic rms (which is often close to null). The mistake stems from the failure of the assumption of the uncertainty to be much less than the estimated quantity, used in the standard propagation of errors.

\section*{ACKNOWLEDGMENTS}

JP has been supported by the Academy of Finland grant 110792. AAZ has been supported by the Academy of Finland exchange grant 112986, the Polish MNiSW grants 1P03D01128 and NN203065933 (2007--2010), and the Polish Astroparticle Network 621/E-78/SN-0068/2007. AI has been supported by the Graduate School in Astronomy and Space Physics, V\"ais\"al\"a foundation and by the Russian Presidential program for support of leading scientific schools (grant NSH-784.2006.2). We thank J. Miko{\l}ajewska for valuable discussion regarding the rotation speed of the companion of Cyg X-1. We are thankful to Guy Pooley for the data from the Ryle telescope. JP and AI acknowledge the support of the 
International Space Science Institute (Bern). JP thanks the Department of Astrophysical Sciences, Princeton University, for hospitality during his visit.
We acknowledge the use of data obtained through the HEASARC online service provided by NASA/GSFC.

\label{lastpage}



\end{document}